\documentclass[twocolumn,twoside,slac_two]{revtex4}
\usepackage{graphicx}
\usepackage{fancyhdr}
\def\NIMA{Nucl.~Instr.~ Methods A}

\def\NPB{Nucl.~Phys.~B~}
\def\PLB{Phys.~Lett.~B~}
\def\PRL{Phys.~Rev.~Lett.~}
\def\PRD{Phys.~Rev.~D~}

\newcommand{\kpee}{K^{\pm}\rightarrow\pi^{\pm}e^{+}e^{-}}
\newcommand{\kppd}{K^{\pm}\rightarrow\pi^{\pm}\pi^{0}_{D}}
\newcommand{\pzd}{\pi^{0}_{D}\rightarrow e^{+}e^{-}\gamma}
\newcommand{\kpgg}{K^{\pm}\rightarrow\pi^{\pm}\gamma\gamma}
\newcommand{\kpeeg}{K^{\pm}\rightarrow\pi^{\pm}e^{+}e^{-}\gamma}
\newcommand{\kppz}{K^{\pm}\rightarrow\pi^{\pm}\pi^{0}}

\newcommand{\ksgg}{K_{S}\rightarrow \gamma\gamma}
\newcommand{\ksee}{K_{S}\rightarrow e^{+}e^{-}}
\newcommand{\ketreg}{K_{L}\rightarrow \pi^{\pm}e^{\mp}\nu(\gamma)}
\newcommand{\ksppz}{K_{S}\rightarrow \pi^{0}\pi^{0}}
\newcommand{\kspp}{K_{S}\rightarrow \pi^{+}\pi^{-}}

\begin{document}

\title{Rare Kaon Decays}


\author{Giuseppina Anzivino}
\thanks{on behalf of the NA48/2 and KLOE Collaborations}
\affiliation{Department of Physics, University of Perugia, Perugia, Italy}
%Collaboration:Cambridge-CERN-Chicago-Dubna-Edinburgh-Ferrara-Firenze-Mainz-Northwestern-Perugia-Pisa-Saclay-Siegen-Torino-Wien}
%

\begin{abstract}
Recent results on charged rare kaon decays from the NA48/2 Collaboration
and on neutral kaon decays from the KLOE Collaboration will be reviewed.
\end{abstract}

\maketitle

\thispagestyle{fancy}


\section{Introduction}

The results presented in this paper are mainly precision tests of
Chiral Perturbation Theory (ChPT), with focus on radiative non-leptonic
kaon decays from the NA48/2 and KLOE Collaborations.
Semileptonic decays ($V_{us}$ related), leptonic and LFV processes and
future rare kaon decays experiments are covered by other talks
in these proceedings.

\section{NA48/2 - Charged kaon decays}
The NA48/2 experiment at CERN has taken data in 2003 and 2004 with the main purpose
of searching for direct CP violation in the decays of charged kaons into three pions.
However, the very high statistics collected, the world's largest, has allowed to 
study many other decays and to measure their Branching Ratios (BR).
In this paper I will fucus on radiative nonleptonic decays, that provide a crucial test
of the ChPT \cite{ecker87} and can give information on the structure of hadronic
interactions at low energy.

\subsection{The NA48/2 experiment}

Two simultaneous focused kaon beams of opposite charge, with a
central momentum of 60 GeV/c and a momentum band of $\pm 3.8 \%$ are
produced by a 400 GeV proton beam impinging on a 40 cm Be target.
The decay volume is a 114 m long vacuum tank; the final
states are reconstructed by a magnetic spectrometer and a liquid
krypton calorimeter (LKr). Charged particles are measured by the
magnetic spectrometer, consisting of four drift chambers and a
dipole magnet located between the second and the third chamber; the
momentum resolution is $\sigma (p)/p=1.02\% \oplus 0.044\%p$ (p in
GeV/c). The magnetic spectrometer is followed by a scintillator
hodoscope consisting of two planes segmented into horizontal and
vertical strips and arranged in four quadrants (charged hodoscope).
The electromagnetic Liquid Krypton calorimeter is an almost
homogeneous ionization chamber with an active volume of $\sim 10~
m^{3}$ and a $27~X_0$ thickness; the energy resolution is
$\sigma(E)/E=0.032/\sqrt{E}\oplus0.09/E\oplus0.0042$ (E in GeV). The
space resolution for single electromagnetic shower can be
parametrized as $\sigma_x=\sigma_y=0.42/\sqrt{E}\oplus0.06$ cm (E in
GeV). At a depth of $\sim 9.5 X_0$ inside the active volume of the
calorimeter, a hodoscope consisting of a plane of scintillating
fibres is installed (neutral hodoscope); the signals from the four
quadrants are used to give a fast trigger.

A more detailed description of the detector can be found
in \cite{batley06}.

\subsection{The $\kpee$ decay}

The $\kpee$ decay is a FCNC process induced at one-loop level in the
Standard Model and highly suppressed by the GIM mechanism.
The dynamics of the decay is completely specified by the invariant function
$W(z)$, where $z=(M_{ee}/M_{K})^{2}$ is a kinematic variable.
Several models have been developed predicting the form factors that
characterize the decay rate and the dilepton invariant mass distribution.
In the present analysis the following parametrizations of the form factors
are considered:
\begin{itemize}
\item Linear: $W(z) = G_{F}M^{2}_{K}f_{0}(1+\delta z)$, with free normalization
and slope $(\delta z)$.
\item Next-to-Leading Order ChPT \cite{d'ambrosio98}:
$W(z) = G_{F}M^{2}_{K}(a_{+}+b_{+} z) + W^{\pi\pi}(z)$
with free parameters $(a_{+},b_{+})$ and an explicitly calculated pion loop term
$W^{\pi\pi}(z)$.
\item A ChPT model developed by a group from Dubna  \cite{dubnickova08} involving
meson form factors $W(z) = W(M_{a},M_{\rho},z)$, with meson masses $(M_{a},M_{\rho})$
treated as free parameters.
\end{itemize}

\begin{figure}[ht]
\centering
\includegraphics[width=80mm]{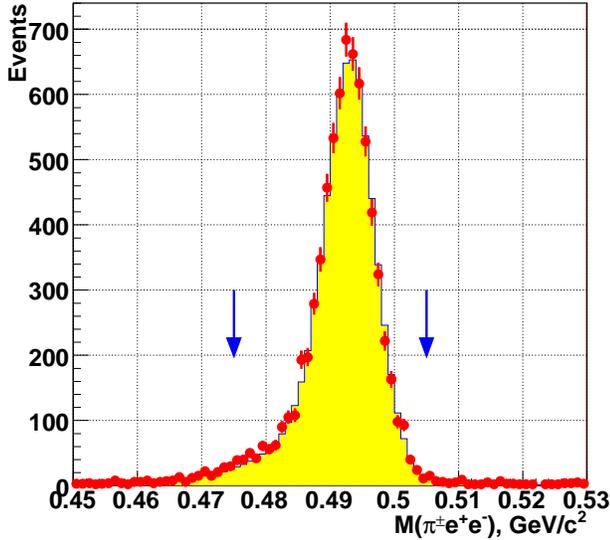}
\caption{Reconstructed spectrum of ($\pi^\pm e^+e^-$) invariant mass:
data (dots) and MC simulation (filled area).} \label{mpiee}
\end{figure}

The aim of the analysis is to extract the form factor parameters in the framework
of each of the above models and to measure the corresponding BR's in
the full kinematic range and, in addition, a model-independent BR in the visible
kinematic range $(z>0.08)$.
The $\kpee$ rate is measured relatively to $\kppd$ (with $\pzd$).
Since the two decays contain the same charged particles in the final state,
common selection criteria have been used, resulting in cancelation of particle
ID inefficiencies at first order.
At the end of the selection, based on the reconstruction of three-track events,
a total of 7146 $\kpee$ candidates with $0.6\%$ background are found in the signal region.
The reconstructed $(\pi^{\pm}e^{+}e^{-})$ invariant mass spectrum is shown in Fig.~\ref{mpiee}.

\begin{figure}[ht]
\centering
\includegraphics[width=80mm]{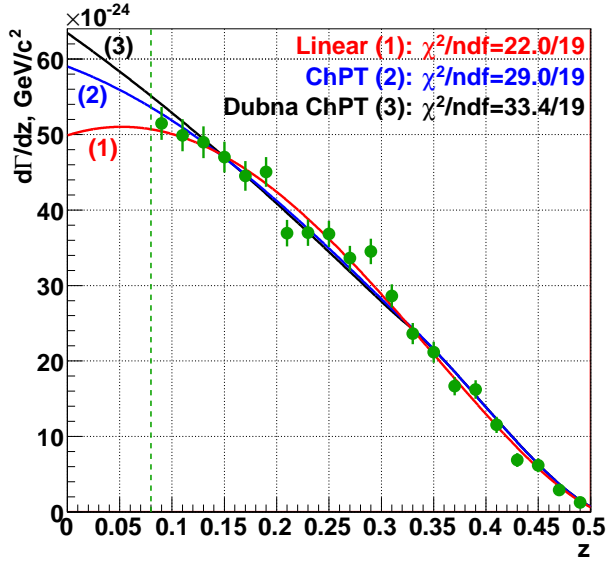}
\caption{The computed $d\Gamma_{\pi e e}/dz$ and the results of the fits according
to the three considered models.} \label{fit}
\end{figure}

The computed values of $d\Gamma_{\pi e e}/dz$ vs $z$ are shown in Fig.~\ref{fit}
with the results of the fits to the three considered models; the measured
parameters and the corresponding BR's are presented in Table~\ref{fit_results}.

\begin{table*}[t]
\begin{center}
\caption{Results of fits to the three models and
the Model-Independent ${\rm BR}(z>0.08)$.} \label{tab:results}
\begin{tabular}{rrrrrrrrrrrr}
\hline
$\delta=\!\!\!$                        &$2.35$  &$\!\pm\!$&$0.15_{\rm stat.}$ &$\!\pm\!$&$0.09_{\rm syst.}$&$\!\pm\!$&$0.00_{\rm ext.}$& $\!\!=\!\!$&$2.35$ &$\!\pm\!$&0.18\\
$f_0=\!\!\!$                           &$0.532$ &$\!\pm\!$&$0.012_{\rm stat.}$&$\!\pm\!$&$0.008_{\rm syst.}$&$\!\pm\!$&$0.007_{\rm ext.}$&$\!\!=\!\!$&$0.532$&$\!\pm\!$&0.016\\
${\rm BR}_1\times10^7=\!\!\!$          &$3.02$  &$\!\pm\!$&$0.04_{\rm stat.}$ &$\!\pm\!$&$0.04_{\rm syst.}$ &$\!\pm\!$&$0.08_{\rm ext.}$ &$\!\!=\!\!$&$3.02$ &$\!\pm\!$&0.10\\
\hline
$a_+=\!\!\!$                           &$-0.579$&$\!\pm\!$&$0.012_{\rm stat.}$&$\!\pm\!$&$0.008_{\rm syst.}$&$\!\pm\!$&$0.007_{\rm ext.}$&$\!\!=\!\!$&$-0.579$&$\!\pm\!$&0.016\\
$b_+=\!\!\!$                           &$-0.798$&$\!\pm\!$&$0.053_{\rm stat.}$&$\!\pm\!$&$0.037_{\rm syst.}$&$\!\pm\!$&$0.017_{\rm ext.}$&$\!\!=\!\!$&$-0.798$&$\!\pm\!$&0.067\\
${\rm BR}_2\times10^7=\!\!\!$          &$3.11$  &$\!\pm\!$&$0.04_{\rm stat.}$ &$\!\pm\!$&$0.04_{\rm syst.}$ &$\!\pm\!$&$0.08_{\rm ext.}$ &$\!\!=\!\!$&$3.11$ &$\!\pm\!$&0.10\\
\hline
$M_a/{\rm GeV}=\!\!\!$                 &$0.965$ &$\!\pm\!$&$0.028_{\rm stat.}$&$\!\pm\!$&$0.018_{\rm syst.}$&$\!\pm\!$&$0.002_{\rm ext.}$&$\!\!=\!\!$&$0.965$&$\!\pm\!$&0.033\\
$M_\rho/{\rm GeV}=\!\!\!$              &$0.711$ &$\!\pm\!$&$0.010_{\rm stat.}$&$\!\pm\!$&$0.007_{\rm syst.}$&$\!\pm\!$&$0.002_{\rm ext.}$&$\!\!=\!\!$&$0.711$&$\!\pm\!$&0.013\\
${\rm BR}_3\times10^7=\!\!\!$          &$3.15$  &$\!\pm\!$&$0.04_{\rm stat.}$ &$\!\pm\!$&$0.04_{\rm syst.}$ &$\!\pm\!$&$0.08_{\rm ext.}$ &$\!\!=\!\!$&$3.15$ &$\!\pm\!$&0.10\\
\hline
${\rm BR_{MI}}\times10^7=\!\!\!$&$2.26$  &$\!\pm\!$&$0.03_{\rm stat.}$ &$\!\pm\!$&$0.03_{\rm syst.}$ &$\!\pm\!$&$0.06_{\rm ext.}$ &$\!\!=\!\!$&$2.26$ &$\!\pm\!$&0.08\\
\hline
\end{tabular}
\label{fit_results}
\end{center}
\end{table*}

Fits to all the three models are of reasonable quality, however the linear
form-factor model leads to the best $\chi^2$ (see Fig.~\ref{fit}).
The data sample is insufficient to distinguish between the models considered.
The BR in the full kinematic range, which includes an
uncertainty due to extrapolation into the inaccessible region $z<0.08$, is
\begin{displaymath}
BR = (3.08\pm0.04_{st.}\pm0.04_{sys.}\pm0.08_{ext.}\pm 0.07_{m.})\times10^{-7}
\end{displaymath}
\vspace{-0.7cm}
\begin{displaymath}
BR(\kpee) = (3.08\pm0.12)\times10^{-7}
\end{displaymath}

This result is in fair agreement with previous measurements. In
particular, comparison to the most precise BNL E865
result~\cite{appel99}, dismissing correlated uncertainties due to
external BR's and model dependence and using the same external input,
reveals a $1.4~\sigma$ difference.
The measurement of $\delta$ is in agreement with the previous
measurements based on $K^+\to\pi^+e^+e^-$~\cite{alliegro92,appel99} and
$K^\pm\to\pi^\pm\mu^+\mu^-$~\cite{ma00} samples, and further
confirms the contradiction of the data to meson dominance
models~\cite{lichard99}. The measured $f_0$, $a_+$ and $b_+$ are in
agreement with the only previous measurement~\cite{appel99}. The
measured parameters $M_a$ and $M_\rho$ are a few \% away from the
nominal masses of the resonances~\cite{pdg}.

\subsection{The $\kpgg$ decay}

\begin{figure}[ht]
\centering
\includegraphics[width=80mm]{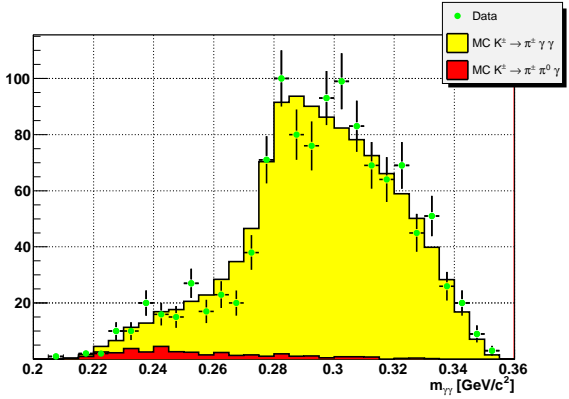}
\caption{Reconstructed invariant $\gamma\gamma$ mass [$GeV/c^2$].} \label{mgg}
\end{figure}

The contributions of the chiral lagrangian to this decay \cite{ecker88}
appear at $\mathcal{O}(p^{4})$, where only the $\Delta I = 1/2$ invariant amplitudes
$A(z)$ and $C(z)$,with $z= M^{2}_{\gamma\gamma}/M^{2}_{K^{\pm}}$, contribute.
The decay rate and the spectrum strongly depend on a single parameter $\hat{c}$
predicted to be positive and $\mathcal{O}(1)$. The invariant $M_{\gamma\gamma}$ distribution
is favored above $2m_{\pi^{+}}$ and exibits a cusp at $2m_{\pi^{+}}$ threshold.
Calculations at $\mathcal{O}(p^{6})$ \cite{d'ambrosio96} show that unitarity correction effects
can increase the BR by $30-40\%$, while vector meson exchange contributions
would be negligible.
The $\kpgg$ decay rate is measured relatively to the $\kppz$ normalization channel;
they have identical particle composition in the final state
and only the cut on the $\gamma\gamma$ invariant mass differs for the two channels.
About 40$\%$ of the total NA48/2 data sample have been analyzed and 1164 candidates
have been found with an estimated background of 3.3 $\%$ which has to be compared
with the only previous experiment \cite{kitching97} that collected 31 events.
This decay and the normalization channel were collected through the neutral trigger
chain, intended for the collection of $K^\pm \to \pi^\pm \pi^0 \pi^0$ and
therefore suffered from a very low trigger efficiency  ($\approx 50\%$).
Elaborate studies were performed in order measure trigger efficiencies and correct for them.
The reconstructed invariant $\gamma\gamma$ mass spectrum in the accessible kinematic
region is shown in Fig.~\ref{mgg}; also shown is the MC expectation assuming
the ChPT prediction of $\mathcal{O}(p^{6})$ \cite{d'ambrosio96}.
The model dependent BR of $\kpgg$ has been measured assuming the validity of
the $\mathcal{O}(p^{6})$ ChPT and $\hat{c}$ = 2.
The following preliminary result has been obtained:
\begin{displaymath}
BR(\kpgg)=(1.07\pm0.04_{stat.}\pm0.08_{syst.})\times10^{-6}
\end{displaymath}
A model independent BR measurement is in preparation, together with the extraction
of $\hat{c}$ from a combined fit of the $m_{\gamma\gamma}$ spectrum shape and the decay rate.

\subsection{The $\kpeeg$ decay}

The $\kpeeg$ decay, observed for the first time by NA48/2,
is similar to the $\kpgg$ with one photon internally converting into a $e^{+}e^{-}$ pair.
After the basic selection criteria, the data were dominated by $\kppd$ decay, where
the $\pi^{0}$ underwent a Dalitz decay $\pzd$.
A number of additional selection criteria had to be applied in order to effectively
suppress the remaining background due to $K^{\pm}$ decays and accidental activity.
The signal region was defined by requiring $480<m_{\pi^{\pm}e^{+}e^{-}\gamma}<505$ [$MeV/c^{2}$].
Since the ChPT predicts only small signal rate and the background increases for low
values of $m_{e^{+}e^{-}\gamma}$, the additional requirement
$m_{e^{+}e^{-}\gamma}$ $>$ 0.26 $GeV/c^{2}$ was applied.
Using the full NA48/2 data sample, 120 $\kpeeg$ (6.1 $\%$ background,
estimated by MC) candidates are found in the accessible kinematic region.

\begin{figure}[h]
\centering
\includegraphics[width=75mm]{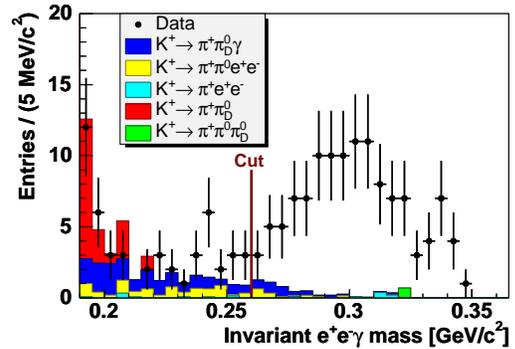}
\caption{Reconstructed invariant $e^{+}e^{-} \gamma$ mass [$GeV/c^2$].} \label{pieeg}
\end{figure}

Fig.~\ref{pieeg} shows the reconstructed spectrum of $e^{+}e^{-}\gamma$, with the MC expectation
for the background contributions.
To determine the BR in a model independent way, a partial branching fraction has been
computed for each $5~MeV/c^{2}$ wide $m_{e^{+}e^{-}\gamma}$ interval $i$:

\begin{displaymath}
BR_{i}(\kpeeg)=\frac{N_{i}^{\pi e e\gamma}-N_{i}^{bkg}}{A_{i}^{\pi e e \gamma}\cdot \epsilon}\times
\frac{1}{\Phi_{K}}
\end{displaymath}

with $N_{i}^{\pi e e\gamma}$ and $N_{i}^{bkg}$ the number of observed signal and
estimated background events, $A_{i}^{\pi e e \gamma}$ the acceptance in bin $i$.
The overall trigger efficiency is $\epsilon$ and $\Phi_{K}$ is the total kaon flux.
By summing over the bins above $m_{e^{+}e^{-}\gamma}$ = 0.26 $GeV/c^{2}$,
the model independent BR is obtained:
\begin{eqnarray}
BR(\kpeeg)(M_{e e \gamma}>0.26 GeV/c^{2}) =  \cr
=(1.19\pm0.12_{stat.}\pm0.04_{syst.})\times10^{-8} \nonumber
\end{eqnarray}

The parameter $\hat{c}$ has also been measured assuming the validity of $\mathcal{O}(p^{6})$
\cite{gabbiani99}. The final results of the analysis have been recently
published \cite{batley08}.

\section{KLOE - Neutral kaon decays}

The results presented in this section are based on data collected with the KLOE detector
at DA$\Phi$NE, the Frascati $e^{+}e^{-}$ collider operated at a center of mass energy of
1020 MeV, the mass of the $\phi$-meson, that is produced almost at rest and decays,
in 34$\%$ of cases, in $K^{0}\overline{K^{0}}$ pairs. The two kaons are always
a $K_{S}K_{L}$ pure pair, so detection of a $K_{L}$ guarentees the presence of a $K_{S}$.
This procedure, called tagging, provides a pure $K_{S}$ beam.

\subsection{The KLOE detector}

The KLOE detector consists of a large cylindrical drift chamber \cite{adinolfi02dc}
of $4~m$ diameter and
$3.3~m$ length operated with a low Z and density gas, surrounded by a lead-scintillating
fiber calorimeter (EMC)\cite{adinolfi02emc}. The chamber provides tracking, measuring
momenta with a resolution
of $\delta(p_{\perp}/p_{\perp})$ of $0.4\%$ at large angle and a vertex reconstruction resolution
of $\sim 3.3~mm$. A superconducting coil around EMC
provides a 0.52 T magnetic field. The EMC is $\sim 15$ $X_{0}$ thick and covers $98\%$
of the solid angle. Energy and time resolutions are $\sigma(E)/E=5.7\%/\sqrt{E}$ and
$\sigma(t)=57~ps/\sqrt{E}\oplus 100~ps$, respectively.

\subsection{The $\ksgg$ decay}

The $\ksgg$ decay is an important test of ChPT. The decay amplitude can be calculated
unambiguously at the leading $\mathcal{O}(p^{4})$ order of the perturbative expansion
with an uncertainty of only few percent \cite{d'ambrosio86},
giving BR ($\ksgg$) = $2.1 \times 10^{-6}$.

In the data sample analyzed, corresponding to an integrated luminosity of
$\int\mathcal{L}dt\sim1.9 fb^{-1}$,
$\sim$ 700$\times10^{6}$ $K_{S}$ decays are identified by $K_{L}$-tagging.
Because of the tagging, no background from $K_{L}\to\gamma\gamma$ decay is expected, the main background
being $\ksppz$, with the two photons undetected, due to geometrical acceptance or non reconstruction
in the calorimeter.
One of the most effective variables against this kind of background is the arrival time of the
photons: a prompt photon is defined as a neutral cluster in the electromagnetic calorimeter satisfying
the condition: $|T-R/c|< min(5~\sigma_{t},2~ns)$, where T is the
time of flight, R is the cluster position with respect to the detector origin of the coordinates
and $\sigma_{t}$ the total time resolution.
A signal-enriched sample is defined by requiring two and not more two prompt photons in the event.
To maximize $\ksppz$ rejection, only clusters with $E>7~MeV$ and $cos\theta < 0.93$ are considered;
the residual background from $K_{S}$ decays other than $2\pi^{0}$ is suppressed by vetoing events
with photons absorbed in a small angle calorimeter.
The event counting is performed using a kinematic fit imposing seven constraints: energy and momentum
conservation, the kaon mass and the two photon velocities.

\begin{figure}[h]
\centering
\includegraphics[width=80mm]{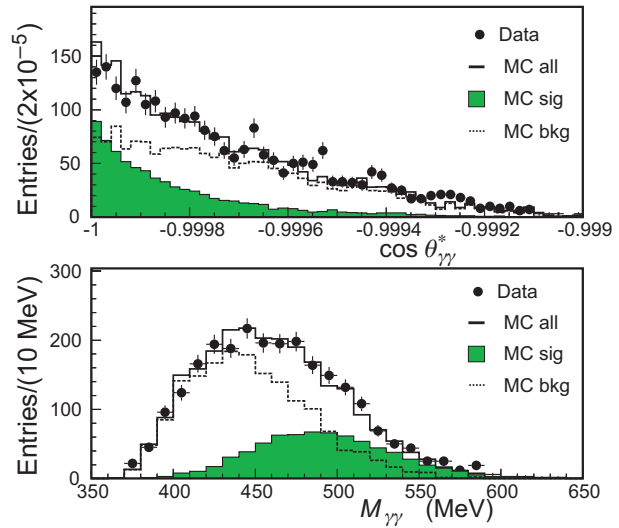}
\caption{Distribution of $\cos\theta^{*}_{\gamma\gamma}$ and $M_{\gamma\gamma}$
for the final sample.} \label{ksgg}
\end{figure}

\begin{figure}[h]
\centering
\includegraphics[width=80mm]{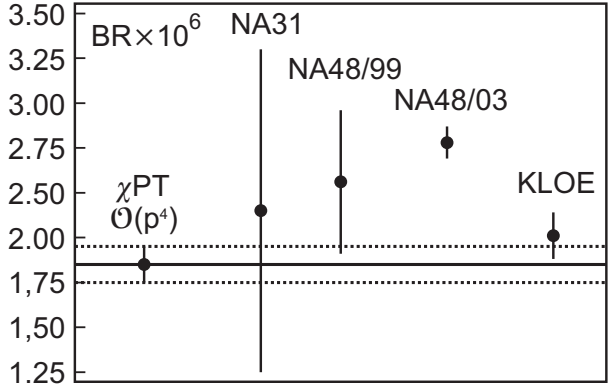}
\caption{Comparison of BR($\ksgg$) measurements and ChPT predictions.} \label{ksggcomp}
\end{figure}

Two other variables with a powerful discrimination against the background are the two photon
invariant mass, $M_{\gamma\gamma}$, and the opening angle between the two photons in the $K_{S}$
center of mass system, $\theta^{*}_{\gamma\gamma}$.
To obtain the number of $\ksgg$ events, a 2 dimensional binned
maximum-likelihood of the final sample distribution in the
$M_{\gamma\gamma}$ and $cos\theta^{*}_{\gamma\gamma}$ variables
is performed using the MC generated signal and background shapes, taking into account data and MC statistics.
The result of the fit is: $N(\gamma\gamma)=711\pm35$, with $\chi^{2}/dof=854/826$, $24.3\%$ CL.
Projections of data and fit are shown in
Fig.~\ref{ksgg}.
The signal $cos\theta^{*}_{\gamma\gamma}$ is peaked  at $cos\theta$=-1, while $M_{\gamma\gamma}$
distribution is gaussian at the $K_{S}$ mass. The background is less peaked at $cos\theta$=-1 and
shows a broader distribution in mass, populating low mass values.
The BR is obtained from N($\ksgg$) using as normalization channel the $\ksppz$ events
recorded in the same data sample by counting the $K_{S}$ tagged events with four prompt photons.
The number of $\ksppz$ events after correcting for selection efficiency is $(190.5\pm0.5)\times10^{6}$.
Using the latest PDG \cite{pdg} value for $BR(\ksppz)= (30.69 \pm 0.05)\%$, the BR($\ksgg$) is obtained
\begin{displaymath}
BR(\ksgg) = (2.26 \pm 0.12_{stat.}\pm0.06_{syst.})\times10^{-6}
\end{displaymath}
This result together with other existing measurements of $BR(\ksgg)$ as well as
the $\mathcal{O}(p^{4})$ ChPT theoretical prediction are shown in Fig.~\ref{ksggcomp}.
There is a $3\sigma$ discrepancy between the
KLOE result and the NA48 latest measurement \cite{NA4803}: the NA48 measurement implies the existence
of a sizable $\mathcal{O}(p^{6})$ counterterm in ChPT, while the KLOE result makes this contribution
practically negligible.

\subsection{The $\ksee$ decay}

\begin{figure}[h]
\centering
\includegraphics[width=70mm]{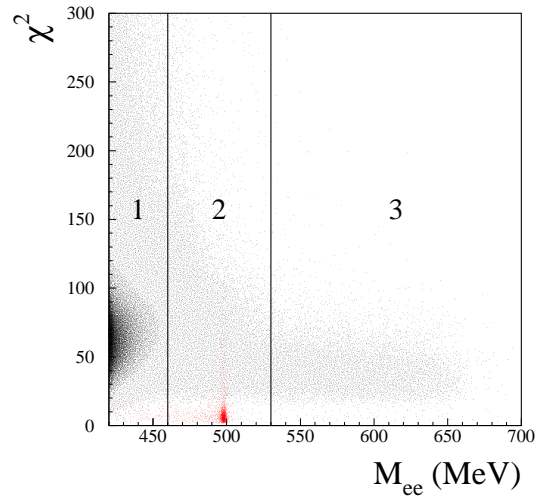}
\caption{Distribution of $\chi^{2}$ as a function of the invariant mass $M_{ee}$:
MC signal (red), background events (black).} \label{ksee}
\end{figure}

The decay $\ksee$, like $K_{L}\to e^{+} e^{-}$ and $K_{L}\to \mu^{+}\mu^{-}$, is a FCNC
process suppressed in the SM and dominated by two photon intermediate state.
Precise theoretical predictions, based on ChPT at $\mathcal{O}(p^{4})$, evaluate the ratio
$\Gamma(\ksee)/\Gamma(\ksgg) = 8 \times 10^{-9}$ with 10 $\%$ uncertainty \cite{ecker91}.
Using the present average for $BR(\ksgg) = (2.71 \pm 0.06) \times 10^{-6}$ \cite{pdg}, the SM prediction
is BR ($\ksee) \sim 10^{-15}$. The best experimental limit has been obtained by the CPLEAR
Collaboration \cite{cplear97}: BR $(\ksee) < 1.4 \times 10^{-7}$ at 90 $\%$ CL.
$K_{S}$ decays are identified by $K_{L}$-tagging and selected by requiring two tracks of
opposite charge with a good vertex. One of the main backgrounds is the $\kspp$ decay,
with two pions misidentified; for these decays the $M_{ee}$ invariant
mass is peaked at low values, so a cut at $M_{ee} > 420~MeV/c$ is effective in rejecting
this type of background. Another important background is the
$\phi \rightarrow\pi^{+}\pi^{-}\pi^{0}$ decay with one prompt photon that simulates
a $K_{L}$ interaction in the EMC and the other photon is not detected.
After preselection, the data sample is reduced to $\sim 10^{6}$ events.
In order to improve the separation between signal and background, a $\chi^{2}$-like variable
is defined, using information from the clusters associated to the candidate electron tracks.
Using the MC signal events, likelihood variables are built based on: the sum and the difference
$\delta t$ of the two tracks, where $\delta t = t_{cl}-L/\beta c$ is evaluated in the
electron hypothesis; the ratio $E/p$ between the cluster energy and the track momentum,
for both charges; the cluster position relative to the extrapolation of the track,
for both charges.
Fig.~\ref{ksee} shows the correlation between $\chi^{2}$ and $M_{ee}$ for MC signal and background.
Two sidebands are defined to check the consistency of MC with data and normalization:
region 1 ($M_{ee}< 460~MeV/c^{2}$), dominated by $\kspp$ events, and region 3
($M_{ee}> 530~MeV/c^{2}$), mostly populated by $\phi \rightarrow\pi^{+}\pi^{-}\pi^{0}$.

A signal box to select $\ksee$ events can be conveniently defined in the $\chi^{2}-M_{ee}$ plane.
Other independent requirements, studied by MC and tuned in the sidebands,
have been applied in order to reduce the background contamination,
before applying the $\chi^{2}-M_{ee}$ selection.
The signal box is chosen with an optimization procedure based only on MC.
The $\chi^{2}$ cut for the signal box definition has been chosen to remove all
MC background events: $\chi^{2} < 70$. The cut on $M_{ee}$ invariant mass is set at
$477 < M_{ee} < 510~MeV/c^{2}$, which rules out all signal events with a radiated
photon with energy greater than $20~MeV$. The signal box selection on data gives
$N_{obs}$ = 0. The upper limit on BR($\ksee$) is evaluated using a bayesian approach
and normalizing to the number of $\kspp$ events. The result is:

\begin{displaymath}
BR(\ksee)\  < \ 9.3\times10^{-9}\ (90 \% \ CL)
\end{displaymath}

This measurement improves by a factor $\sim$ 15 the CPLEAR \cite{cplear97} result and
includes for the first time radiative corrections.

\subsection{The $\ketreg$ decay}

\begin{figure}[ht]
\centering
\includegraphics[width=70mm]{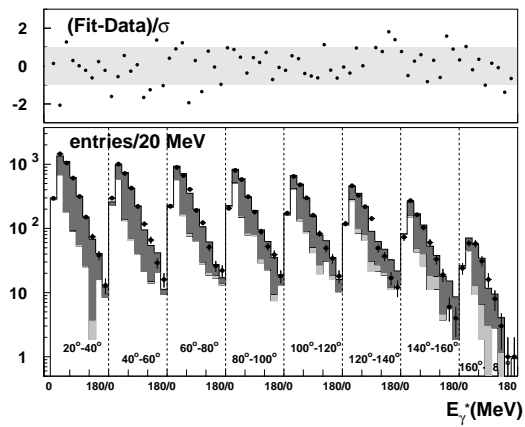}
\caption{Fit and fit residuals in bins of $E^{*}_{\gamma}$.} \label{ke3g}
\end{figure}

The study of radiative $K_{L}$ decays offers the possibility to obtain information on the kaon structure
and to test ChPT. Two different components contribute to the photon emission: the inner
bremsstrahlung (IB) and the direct emission (DE). In the $K^{0}_{e3 \gamma}$ decay the IB component
is much larger than the DE and the interference terms.
It is customary to apply standard cuts to allow comparison among results: $E^{*}_{\gamma}$ $>$ 30 MeV
and $\theta^{*}_{\gamma}$ $>$ 20° ($E^{*}_{\gamma}$ and $\theta^{*}_{\gamma}$ energy
and photon angle w.r.t. the lepton in the kaon rest frame, respectively).
The ratio R is defined:
\begin{displaymath}
R=\frac{\Gamma(K^{0}_{e 3\gamma};E^{*}_{\gamma} > 30~MeV,\theta^{*}_{\gamma} > 20°)}{\Gamma(K^{0}_{e 3(\gamma)})}
\end{displaymath}
Theoretical prediction for R range between 0.95 $\times 10^{-2}$ and $0.97 \times 10^{-2}$ \cite{gasser05}.
The decay amplitude can be written as
\begin{displaymath}
\frac{d\Gamma}{dE^{*}_{\gamma}}\simeq\frac{d\Gamma_{IB}}{dE^{*}_{\gamma}}+\langle X \rangle f(E^{*}_{\gamma})
\end{displaymath}
where the second term is the SD contribution. All the information on the SD term is contained
in the effective strength $\langle X \rangle$.
Candidate $K_{L}$ events are tagged by the presence of a $\kspp$ decay.
To count $K^{0}_{e3 \gamma}$ signal events, a two-dimensional fit in the variables
$E^{*}_{\gamma}, \theta^{*}_{\gamma}$ is performed; this allows to measure
both R and $\langle X \rangle$. The results of the fit are shown in Fig.~\ref{ke3g}.
\begin{figure}[h]
\centering
\includegraphics[width=70mm]{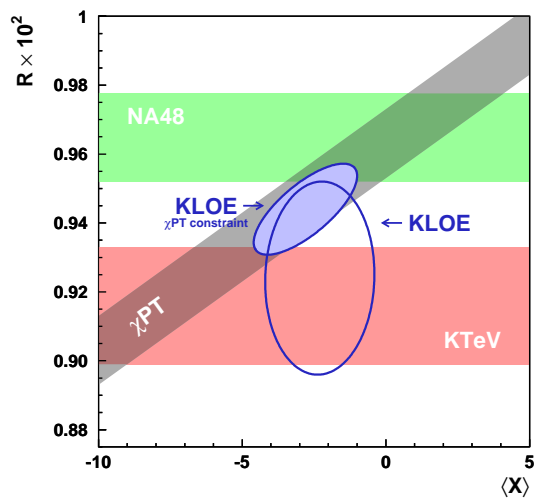}
\caption{Comparison of $R$ and $\langle X \rangle$ results (see text).} \label{compke3}
\end{figure}

Taking into account all systematics, the measurement of R and $\langle X \rangle$ yields
%\begin{eqnarray}
%R &=& (924 \pm 23_{stat} \pm 16_{syst})\times 10^{-5} \cr
%\langle X \rangle &=& -2.3 \pm 1.3_{stat} \pm 1.4_{syst} \nonumber
%\end{eqnarray}

\begin{displaymath}
R = (924 \pm 23_{stat} \pm 16_{syst})\times 10^{-5}
\end{displaymath}
\vspace{-0.6cm}
\begin{displaymath}
\langle X \rangle = -2.3 \pm 1.3_{stat} \pm 1.4_{syst}
\end{displaymath}

The measured value of $\langle X \rangle$ is in agreement with $\mathcal{O}(p^{6})$ evaluation in \cite{gasser05}.
The presence of DE contribution reduces the value of R of about 1 $\%$.
Fig.~\ref{compke3} shows the comparison of this results with previous measurements.
The present accuracy is not sufficient to solve the experimental discrepancy between
NA48 \cite{lai05} and KTeV \cite{alexopolous05}.

%\begin{thebibliography}{9}   % Use for  1-9  references

\end{document}